# Generative Replica-Exchange: A Flow-based Framework for Accelerating Replica Exchange Simulations

Shengjie Huang[1], Sijie Yang[1], Jianqiao Yi[1], Rui Zheng[1], Haocong Liao[1], Muzammal Hussain[2,3], Yaoquan Tu[4], Xiaoyun Lu[1,*], Yang Zhou[1,*]

[1] State Key Laboratory of Bioactive Molecules and Druggability Assessment, International Cooperative Laboratory of Traditional Chinese Medicine Modernization and Innovative Drug Discovery of Chinese Ministry of Education (MOE), School of Pharmacy, Jinan University, #855 Xingye Avenue, Guangzhou, 510632, China

[2] Department of Biochemistry and Molecular Pharmacology, NYU Grossman School of Medicine, New York, NY 10016, USA

[3] Howard Hughes Medical Institute, NYU Grossman School of Medicine, New York, NY 10016, USA

[4] Department of Theoretical Chemistry and Biology, KTH Royal Institute of Technology, Stockholm 114 28, Sweden

**Abstract**

Replica exchange (REX) is one of the most widely used enhanced sampling methodologies, yet its efficiency is limited by the requirement for a large number of intermediate temperature replicas. Here we present Generative Replica Exchange (GREX), which integrates deep generative models into the REX framework to eliminate this temperature ladder. Drawing inspiration from reservoir replica exchange (res-REX), GREX utilizes trained normalizing flows to generate high-temperature configurations on demand and map them directly to the target distribution using the potential energy as a constraint, without requiring target-temperature training data. This approach reduces production simulations to a single replica at the target temperature while maintaining thermodynamic rigor through Metropolis exchange acceptance. We validate GREX on three benchmark systems of increasing complexity, highlighting its superior efficiency and practical applicability for molecular simulations.

# 1. Introduction

Molecular dynamics (MD) is a powerful computational technique for simulating complex phenomena at the atomic level[1]. Recent advances in computing hardware and software have significantly extended the accessible simulation timescales, ranging from nanoseconds to microseconds, thereby enabling researchers to investigate dynamic processes, such as small peptide folding or ligand binding[2,3]. Despite these advancements, achieving efficient sampling of the equilibrium distribution remains a long-standing challenge in molecular simulations. This difficulty arises because systems often become trapped in local energy minima, which limits their ability to explore configurational space in short simulation times. Furthermore, achieving sufficient sampling on long timescales remains computationally demanding, especially for large systems or rare-event processes.

To address these challenges, a variety of enhanced sampling techniques have been developed to facilitate the traversal of energy barriers and accelerate the exploration of configurational space. These methodologies are generally categorized into two frameworks: collective variable (CV)-based and CV-free approaches. CV-based methods, such as metadynamics[4,5] and umbrella sampling[6], rely on the selection of a few critical degrees of freedom to bias the simulation and promote sampling along relevant coordinates. However, the efficacy of these methods is heavily dependent on the "quality" of the selected CVs[7]. CV-free strategies, such as replica exchange molecular dynamics (REX)[8,9], facilitate barrier crossing without relying on predefined

CVs. By allowing replicas to swap configurations between neighboring distributions, the system can effectively escape local minima. Driven by the need for better scalability, several variants of REX have emerged such as Hamiltonian REX (HREX)[10,11], Replica Exchange Solute Tempering (REST)[12], and reservoir replica exchange (res-REX)[13]. In particular, in res-REX a reservoir of configurations is pre-generated at a high temperature, and exchange attempts are performed between the reservoir and replicas. This method greatly enhances the convergence rate and has been successfully used for a variety of small molecules and peptides such as leucine tripeptide[14], Trpzip2[15], Aβ$_{21-30}$ peptide[16]. Nevertheless, conventional REX and its variants still require a large number of intermediate temperature replicas to maintain adequate exchange acceptance probabilities, resulting in significant computational overhead that grows with system size and complexity.

In this work, we present Generative replica exchange (GREX), an enhanced sampling framework that integrates normalizing flows into the conventional replica exchange framework to eliminate the need for a dense temperature ladder, reducing production simulation to a single replica at the target temperature. To demonstrate its effectiveness and generality, we apply GREX to three benchmark systems of increasing complexity: a double-well potential, alanine dipeptide in explicit water, and the 10-residue mini-protein chignolin. GREX achieves convergence speedups of 5- to 10-fold relative to conventional REX while maintaining thermodynamic accuracy, with computational advantages that grow with system complexity.

## 2. Methods

**Reservoir Replica Exchange Simulations (res-REX)**

We briefly summarize the key aspects of REX and res-REX as they are related to the present study. In standard REX, $N$ replicas ($R^1$ to $R^N$) at different temperatures are simulated simultaneously and independently for a chosen number of MD steps. Exchanges between a pair of replicas $R^i$ and $R^j$ (typically neighboring replicas where $j = i + 1$) are attempted periodically. The acceptance probability for an exchange is defined by the Metropolis criterion,

$$P(R^i \leftrightarrow R^j) = \min\left(1, \exp\left[\left(\frac{1}{k_B T_i} - \frac{1}{k_B T_j}\right)(U_i - U_j)\right]\right), \tag{1}$$

where $T_i$ and $T_j$ are the temperatures of $R^i$ and $R^j$ and $U_i$ and $U_j$ are the potential energies of the configurations at $R^i$ and $R^j$, respectively. If the exchange is accepted, then the bath temperatures of these replicas will be swapped. Otherwise, if the exchange is rejected, then each replica will continue on its current trajectory with the same bath.

Reservoir Replica Exchange Simulations (res-REX) is a variant of standard REX. In this method, the highest temperature replica ($R^N$) is replaced with a reservoir ($R^r$), which is a set of structures previously generated from MD simulations performed at the high temperature $T^N$. The exchange attempt is made between the structure of $R^{N-1}$ and a randomly selected structure from the reservoir. If the exchange is accepted, the coordinates from $R^r$ are sent to replica $R^{N-1}$ and the chosen reservoir structure is left in the reservoir, as it is assumed that the reservoir constitutes a complete representation of the ensemble and that the inclusion of the new coordinates will have a negligible effect on the reservoir. Concurrently, standard REX simulations were used for each of the lower temperatures (replicas $R^1$ to $R^{N-1}$, $T^1$ to $T^{N-1}$) and exchanges are attempted on the basis of the same criterion as used for REX (eq 1). The standard REX process of simulation across the temperature ladder provide exploration/refinement of the basins present in the reservoir and also reweights the probability of observing these structures at different temperatures. The exchanges between $R^r$ and $R^{N-1}$ and the REX for $R^1$ to $R^{N-1}$ are repeated multiple times and thermally reweighted many times during the simulation, resulting in converged Boltzmann-weighted ensembles at all temperatures.

**GREX framework**

GREX extends the res-REX concept by replacing the static high-temperature reservoir with a Generator Flow (GF) that generate configurations reproducing the high-temperature distribution, and introducing a Converter Flow (CF) that maps these configurations directly to the target temperature. Together, they eliminate the intermediate replica ladder and reducing the simulation to a single target-temperature replica.

The GREX workflow has two main stages, as illustrated in Figure 1. In Stage 1 (training), a short MD simulation is performed at the high temperature $T^N$ to collect configurations. These samples are used to train the GF to reproduce the high-temperature distribution $p_{x_h}$, while the CF is trained with the potential energy function as a physical constraint to learn the mapping from $p_{x_h}$ to the target distribution $p_{x_l}$. In Stage 2 (production), the GF generates high temperature configurations, which are then mapped to the target temperature via the CF. Exchange attempts between the mapped configurations and the ongoing MD trajectory are evaluated using the Metropolis criterion (eq 1), ensuring thermodynamic consistency at the target temperature throughout the simulation.

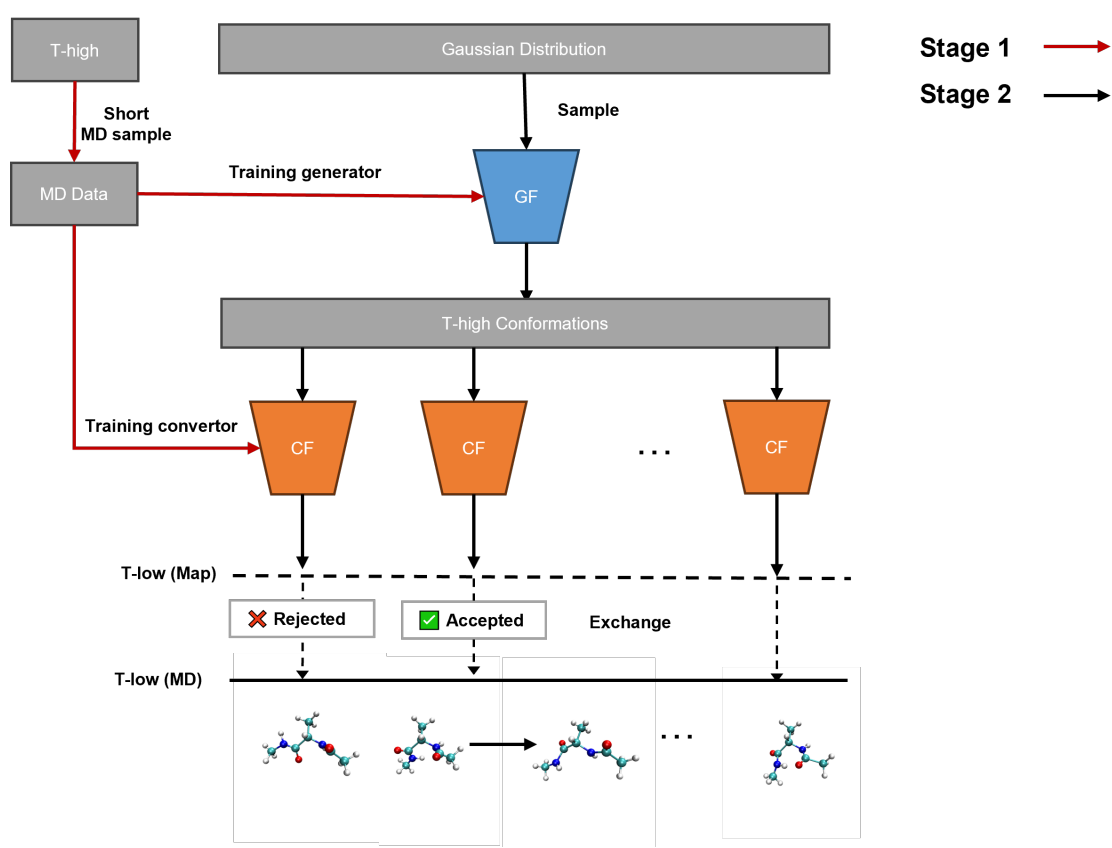

Figure 1. The GREX workflow. The blue block represents the generator flow (GF), and orange blocks represent the converter flow (CF). Red arrows indicate State 1, while black arrows indicate State 2.

Both the GF and CF are implemented as normalizing flows (as shown in Figure 2), which are trainable invertible neural networks ($f$ and $f^{-1}$) that enable exact and tractable computation of probability densities.

**Generator Flow (GF)**

The GF is a normalizing flow that learns to transform samples from a prior Gaussian distribution $p_z(z)$ into configurations distributed according to the high temperature Boltzmann distribution $p_{x_h}(x_h)$. Once trained, the GF serves as an on-demand generator of high-temperature configurations, replacing the role of the pre-generated reservoir in res-REX.

The GF implements this mapping through a series of reversible coordinate transformations[17–19], expressed as follows:

$$x_h = f(z; \theta) \tag{2}$$

$$z = f^{-1}(x_h; \theta) \tag{3}$$

where, $x_h$ and $z$ represent samples from the high-temperature distribution and the prior distribution, respectively, and $\theta$ represents the trainable parameters of GF. Each transformation is associated with a Jacobian matrix:

$$J_{x_h \to z}(x_h; \theta) = [\frac{\partial f^{-1}(x_h; \theta)}{\partial x_{h1}}, \dots, \frac{\partial f^{-1}(x_h; \theta)}{\partial x_{hn}}] \tag{4}$$

$$J_{z \to x_h}(z; \theta) = [\frac{\partial f(z; \theta)}{\partial z_1}, \dots, \frac{\partial f(z; \theta)}{\partial z_n}] \tag{5}$$

The absolute value of the Jacobian determinant, $|\det J_{z \to x_h}(z; \theta)|$ quantifies the degree of volume expansion or contraction induced by the transformation. Owing to the invertibility of the mapping, probability densities can be transformed between different spaces as

$$q_{x_h}(x_h) = p_z(f^{-1}(x_h; \theta))|\det J_{x_h \to z}(x_h; \theta)| \tag{6}$$

$$q_z(z) = p_{x_h}(f(z; \theta))|\det J_{z \to x_h}(z; \theta)| \tag{7}$$

where $q$ denotes the distributions generated by the neural network in the corresponding spaces, which differ from the target distributions $p$.

The GF is built upon the affine coupling layers of the RealNVP algorithm[20] proposed by Dinh et al, following the design of Boltzmann generators[21]. Specifically, the input variables are partitioned into two channels, denoted as $x = (x_1, x_2)$ and $z = (z_1, z_2)$. In each affine coupling layer, one part remains unchanged, while the other part is transformed using a scaling network $S$ and a translation network $T$, both implemented as non-invertible neural networks. The corresponding transformations are given by

$$f^{-1}((x_1, x_2); \theta): \begin{cases} z_1 = x_1, \\ z_2 = x_2 \odot \exp(S(x_1)) + T(x_1), \end{cases} \tag{8}$$

$$f((z_1, z_2); \theta): \begin{cases} x_1 = z_1, \\ x_2 = (z_2 - T(x_1)) \odot \exp(-S(z_1)), \end{cases} \tag{9}$$

where $\odot$ denotes element-wise multiplication.

To reconstruct the high-temperature distribution $p_{x_h}(x_h)$, the GF is trained using a "training by example" strategy[21]. High-temperature trajectory data from Stage 1 serve as the training data, and the network parameters $\theta$ are optimized by minimizing the negative log-likelihood (NLL) loss function[22,23]:

$$L_{NLL} = -\mathbb{E}_{x_h \sim p_{x_h}(x_h)}\left[\log p_z(f^{-1}(x_h; \theta))|\det J_{x_h \to z}(x_h; \theta)|\right] \quad (10)$$

**Converter Flow (CF)**

The CF is a normalizing flow that map the high-temperature distribution $p_{x_h}(x_h)$ directly to the target low-temperature distribution $p_{x_l}(x_l)$, eliminating the need for intermediate-temperature replicas. Following the LREX method introduced by Invernizzi et al.[24], the CF achieves this distribution transformation through:

$$q_{x_l}(x_l) = p_{x_h}(f^{-1}(x_l; \phi))|\det J_{x_l \to x_h}(x_l; \phi)| \quad (11)$$

$$q_{x_h}(x_h) = p_{x_l}(f(x_h; \phi))|\det J_{x_h \to x_l}(x_h; \phi)| \quad (12)$$

where $x_l$ represents samples from the low-temperature distribution, and $\phi$ denotes the trainable parameters of CF.

Unlike the GF, which is trained using a "training by example" strategy, the CF faces the major challenge that samples from the target low-temperature distribution $p_{x_l}(x_l)$ are generally unavailable. The CF therefore adopts a "training by energy" strategy[24], in which the system's potential energy function $U(x)$ is incorporated as a physical constraint. Specifically, the network parameters $\phi$ are optimized by the Kullback–Leibler divergence (KLD) between the generated and target distributions[24]:

$$L_{KLD} = \mathbb{E}_{x_h \sim p_{x_h}(x_h)}[\frac{1}{k_B T_l}U(f(x_h;\phi)) - \frac{1}{k_B T_h}U(x_h)$$
$$- \log|\det J_{x_h \to x_l}(x_h;\phi)|] \qquad (13)$$

where $k_B$ is the Boltzmann constant.

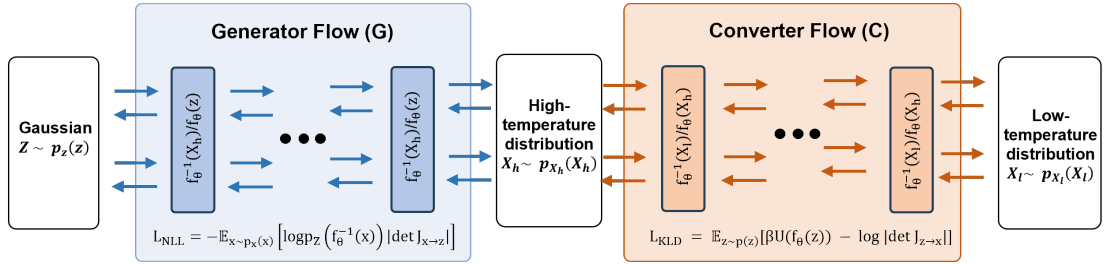

Figure 2. Architecture of Generator Flow (G) and Converter Flow (C)

**Alanine dipeptide and Chignolin**

The structure of alanine dipeptide was modelled using AmberTools23[25] and the structure of chignolin (PDB ID: 1UAO) was obtained from the RCSB Protein Data Bank[26]. Both systems were simulated using OpenMM[27], with the AMBER ff14SB force field[28] for the protein and the TIP3P[29] model for the solvent. Each system was solvated in a cubic simulation box with a minimum distance of 1.2 nm between the systems and the box boundaries. After solvation, an energy minimization process was carried out to remove the atomic clashes and optimize the geometry of all molecules. The simulations were run in the constant number (N), pressure (P), and temperature (T) (NPT) ensemble at 300 K and 1 bar using MonteCarlo algorithm. The Particle Mesh Ewald (PME)[30] method was used for calculating long-range electrostatic interactions. The LangevinMiddleIntegrator[31] method was used with a time step of 2 fs. Following the simulations, the MDtraj package[32] was used to extract protein atoms, and the trajectories were aligned based on CA atom pairs to generate the dataset. For alanine dipeptide, the backbone dihedral angles Φ and Ψ were calculated. For chignolin, RMSD of simulation frames relative to the PDB native structure was calculated for the protein Cα atoms with the terminal residues Gly1 and Gly10 excluded.

## 3. Results

To demonstrate the efficiency and capabilities of GREX, we apply it to three model systems of increasing complexity: particle in double-well potential, alanine dipeptide, and chignolin.

### 3.1 Double-well potential

To validate the algorithmic correctness and the sampling efficiency of GREX, we first applied it to a particle moving with Langevin dynamics in a double-well potential[33] (see SI for details). The target temperature was set to $T = 1$, at which transitions between the two basins are exceedingly rare in cMD simulations. A separate cMD simulation at a higher temperature ($T = 5$), where barrier crossing occurs much more frequently (Figure S1A), was performed to generate training data for both the generator and converter flows.

As illustrated in Figure 3A, at the target temperature, the particle remains trapped in the left basin ($x < 0$) throughout the 20 ns cMD simulation. In contrast, frequent transitions between the two basins were observed in both GREX and REX (Figures 3B and 3C). The one-dimensional marginal probability distributions along the x-coordinate demonstrate that both methods successfully recover the free energy difference between the two basins (ΔG = 2.5 kcal/mol) in excellent agreement with the analytical solution (Figure S1B), confirming that GREX reproduces the correct equilibrium free energy landscape.

Next, to evaluate the efficiency of GREX in comparison with REX, we followed the protocol of Invernizzi et al. by extending the double-well potential system to N dimensions[24]. In conventional REX, the number of intermediate temperature replicas required to maintain adequate exchange acceptance ratios scales with system size, leading to a dramatic increase in computational cost as dimensionality grows. In contrast, GREX requires sampling only at the target temperature, with the generator and converter flows establish the connection to the high-temperature distribution. To quantify the resulting difference in scalability, we measured the computational time required to reach convergence for systems of varying dimensionality, using the basin occupation fraction (the fraction of time with x > 0) as a convergence indicator. Figures 3D–3H present the time evolution of this quantity for systems with different dimensions. For small systems (e.g., N = 2), GREX and REX perform comparably, with both methods requiring approximately 50 seconds to approach equilibrium. As the system size increases, however, the convergence time required for REX increases dramatically, reaching approximately 300 s, 600 s, and 2000 s for N = 64, 512, and 1024, respectively. In contrast, GREX simulations reach equilibrium with minimal dependence on system size, maintaining convergence times around 200 s for N ≤ 1024, with only modest increases observed for larger systems (Figure 3I).

We also examined the effect of system size on exchange acceptance ratios. In REX, maintaining reasonable acceptance ratios necessitates a proportional increase in the number of intermediate replicas as system size grows. Conversely, GREX consistently maintains high acceptance ratios (> 20%) across all system sizes without requiring additional replicas (Figure 3J), demonstrating superior scalability. Taken together, these results establish that GREX offers substantial computational advantages over conventional REX, particularly for high-dimensional systems where the replica ladder approach becomes prohibitively expensive.

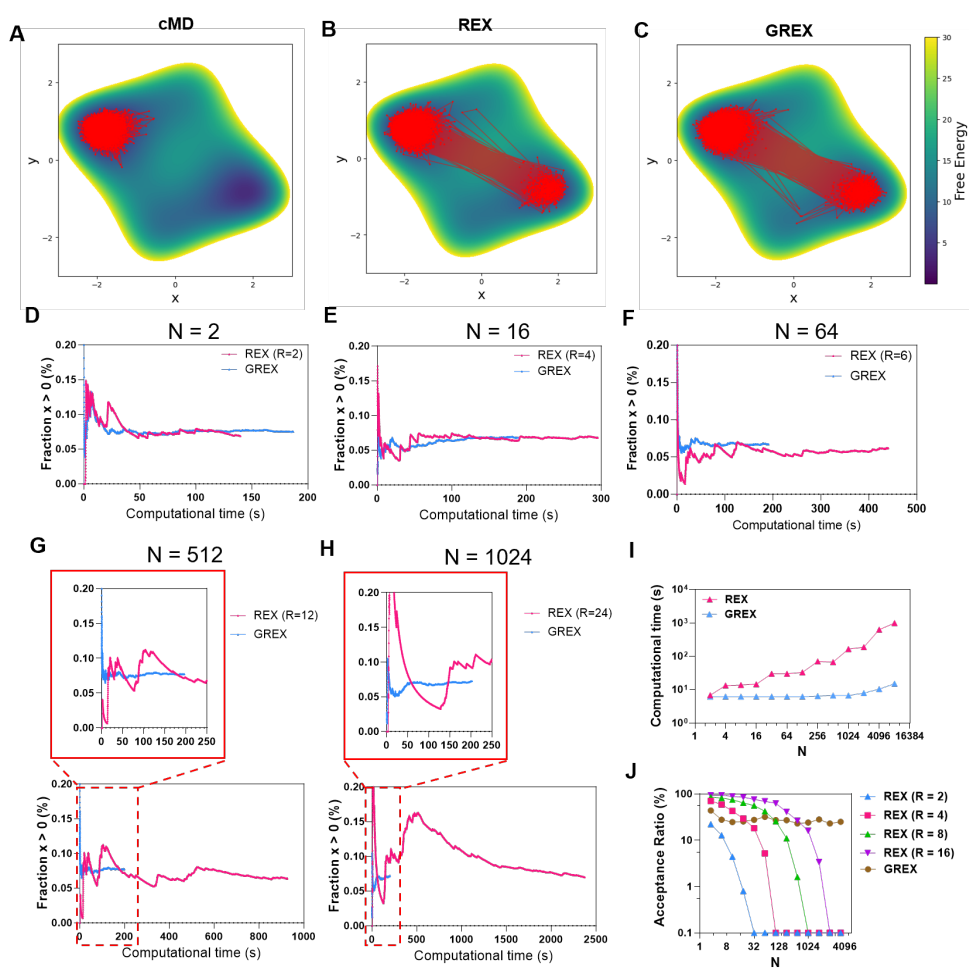

Figure 3. Evaluation of the double-well potential model. (A-C) The free energy surface and representative trajectories obtained from cMD, REX and GREX simulations at the target temperature ($T = 1$). (D-H) Time evolution of the basin occupation fraction (defined as the fraction of configurations with $x > 0$) as a function of computational time for $N$-dimensional systems with increasing dimensionality ($N = 2, 16, 64, 512,$ and $1024$). (I) Computational time required for convergence as a function of system dimensionality N. (J) Exchange acceptance ratios increase with N for REX simulations with different numbers of replicas ($R$) and for GREX.

### 3.2 Alanine Dipeptide

We next applied the method to alanine dipeptide in explicit water, a widely used benchmark system whose conformational landscape is well-characterized by the two backbone dihedral angles $\Phi$ and $\Psi$ (Figure 4A) [34,35]. Five independent 2 μs cMD simulations at 300 K were used to generate a reference free energy surface against which all methods were compared. These method included conventional REX using 32 replicas spanning 300-1000 K, res-REX using 31 replicas supplemented with a 1000 K reservoir constructed from 20 ns cMD data, and GREX using a single 300 K replica with generative models trained on the same 20 ns high-temperature trajectory.

All three methods reproduced the reference FES obtained from cMD simulations (Figure 4A and S2), confirming that GREX achieves results comparable to replica-based approaches despite using only a single production replica. To quantify sampling efficiency, we monitored the convergence of conformational basins {C5, $P_{II}$, $\alpha_R$, and α'} as a function of computational time. For the low-energy basins {C5, P$_{II}$, and α$_R$}, GREX converged within approximately 5000 s, roughly half the time required by both res-REX and REX (~10000 s; Figure S3). The advantage was even more pronounced for the high-energy α' basin: GREX again converged within ~5000 s, whereas both res-REX and REX required ~30000 s, corresponding to an approximately sixfold difference (Figure 4C).

Because GREX training depends entirely on high-temperature simulation data, the amount of training data required is a key practical consideration. To investigate this, we trained GREX models on 1000 K trajectories ranging from 0.01 ns to 20 ns and performed 100 ns production simulations for each case. With only 0.01 ns of training data, GREX failed to recover the reference free energy landscape, producing a distribution that was entirely absent in the region Φ > 0° (Figure 4D). This failure arise because 0.01 ns trajectory does not sample that region of conformational space. Once the training trajectory covered the relevant conformational regions, however, GREX accurately reproduced the 300 K reference distribution. Notably, this did not require the high-temperature FES itself to be converged. Models trained on 0.05 ns and 0.5 ns trajectories, whose high-temperature FES remain far from the converged reference, still reproduce the 300 K reference landscape with reasonable accuracy (Figures 4E to 4J). With 2 ns of training data, GREX already matched the accuracy of both res-REX and GREX model trained on 20 ns of data (Figures 4G and 4H). Accounting for this training cost, the total computational expense of GREX was approximately ~16000 s (2 ns training plus 100 ns single-replica production), compared with approximately ~157000 s for conventional REX (32 replicas × 100 ns), yielding an overall speedup of roughly 10-fold.

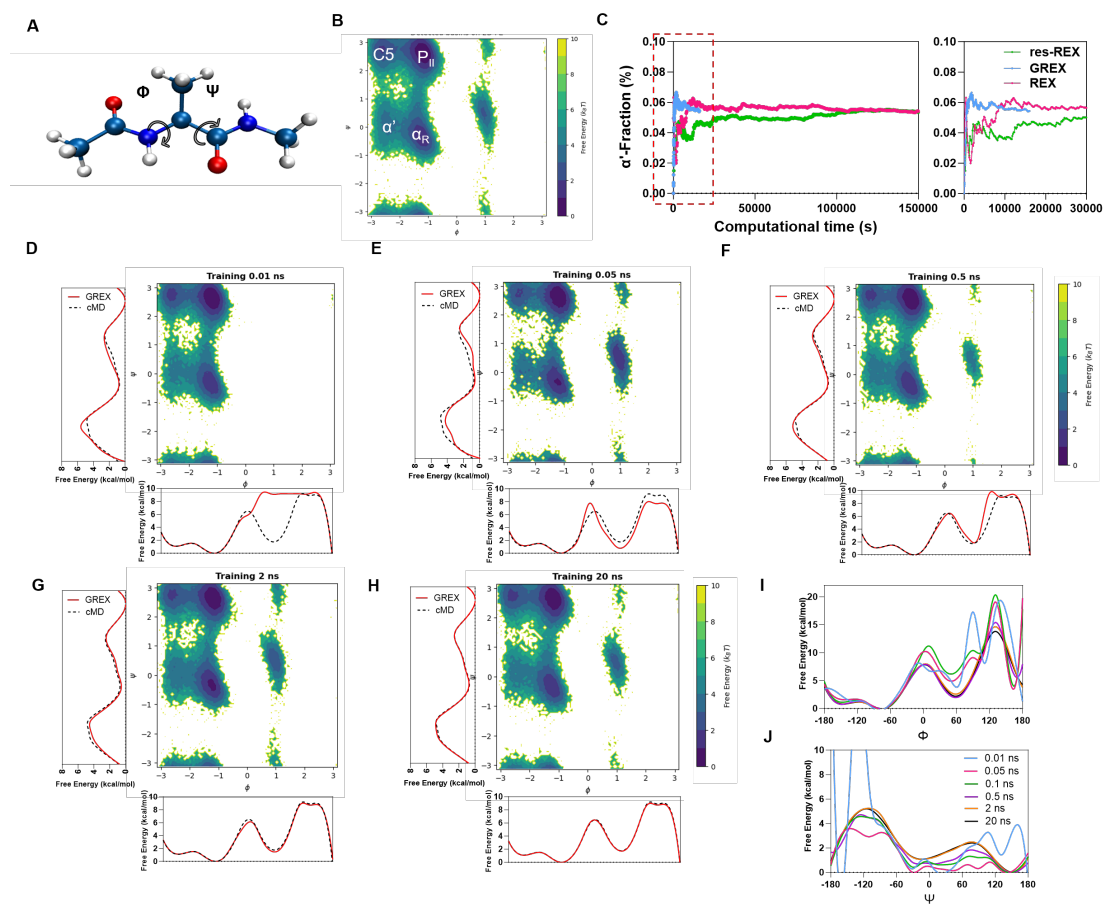

Figure 4. Evaluation of the alanine dipeptide model. (A) Illustration of the backbone dihedral angles Φ and Ψ of the alanine dipeptide. (B) Reference free energy surface at 300 K as a function of Φ and Ψ, obtained from cMD simulations. Major conformational basins {C5, P$_{II}$, α$_R$, and α'} are labeled. (C) Convergence of the α' basin population as a function of computational time for GREX, res-REX, and conventional REX. The dashed box highlights the early-time regime. (D-H) Free energy surfaces reconstructed from GREX simulations at 300 K using high-temperature (1000 K) cMD training data of different lengths: (D) 0.01 ns, (E) 0.05 ns, (F) 0.5 ns, (G) 2 ns, and (H) 20 ns. (I) and

(J) One-dimensional free energy profiles along Φ and Ψ obtained from high-temperature cMD simulations with different trajectory lengths.

**3.3. Chignolin**

To further assess the applicability of GREX to more complex biomolecular systems, we examined chignolin, a 10-residue mini-protein (GYDPETGTWG) that folds into a stable β-hairpin structure. Chignolin has been extensively characterized both experimentally and computationally, making it an ideal benchmark for enhanced sampling methods[36–39]. We generated training data from a 50 ns MD simulation at 500 K, with the target temperature set to 300 K. Three independent 10 μs cMD simulations at 300 K served as reference data, and REX simulations with 24 replicas (300–500 K) were used for comparison.

We first characterized the overall conformational landscape by computing the two-dimensional FES as a function of backbone RMSD and radius of gyration (Rg). The reference FES from 10 μs cMD simulations (Figure 5A) reveals a basin at (RMSD ≈ 1.2 Å, Rg ≈ 5.8 Å). Both GREX and REX successfully reproduced this landscape (Figures 5B and 5C), confirming that both methods recover the correct equilibrium structural ensemble at the target temperature.

We then assessed folding/unfolding transitions using backbone Cα RMSD relative to the experimental structure. In the 10 μs cMD simulations, chignolin exhibited characteristic two-state behavior, remaining in either the folded (RMSD < 2.0 Å) or unfolded states (RMSD > 4.0 Å) for hundreds of nanoseconds to microseconds (Figure

5D). In contrast, both GREX and REX sampled frequent folding/unfolding transitions within the 100 ns simulations (Figure 5E–F). Notably, GREX also accessed near-native conformations with a minimum RMSD of 3.0 Å (Figure 5G), prompting a quantitative assessment of folding accuracy.

Therefore, to evaluate the accuracy of GREX, we estimate the Gibbs free energy difference ΔG between the unfolded and the folded states:

$$\Delta G = -k_B T \ln\left(\frac{p_f}{1 - p_f}\right) \tag{14}$$

where $k_B$ is the Boltzmann constant, temperature $T$ is set to 300 K, $p_f$ is the probabilities of chignolin's conformations in the folded conformations. Honda et al.[40] measured the folding free energy ΔG using circular dichroism and NMR techniques obtaining values between 0.26 and 0.45 kcal/mol at 298 K. GREX estimate of $\Delta G_f$ = 0.47 kcal/mol that is in close agreement with the experimental range (Figure 5H, dashed line), confirming that GREX accurately captures the thermodynamic stability of chignolin at the target temperature.

We further compared convergence efficiency by monitor $\Delta G_f$ over simulation time for all three methods (Figure 5H). In cMD simulation, $\Delta G_f$ converges after approximately 600000 s (10 μs) to a value of about 0.43 kcal/mol (Figure 5H). In REX, convergence reached 0.73 kcal/mol after 200000 s (24 replicas × 100 ns), significantly outside the experimental range, indicating incomplete convergence despite the higher computational cost. GREX, in contrast, achieved convergence to 0.47 kcal/mol within only 20,000 s (100 ns), consistent with both the experimental value and the long-time cMD simulation data. These results demonstrate that GREX achieves superior convergence efficiency related to REX while maintaining accuracy comparable to long-timescale cMD.

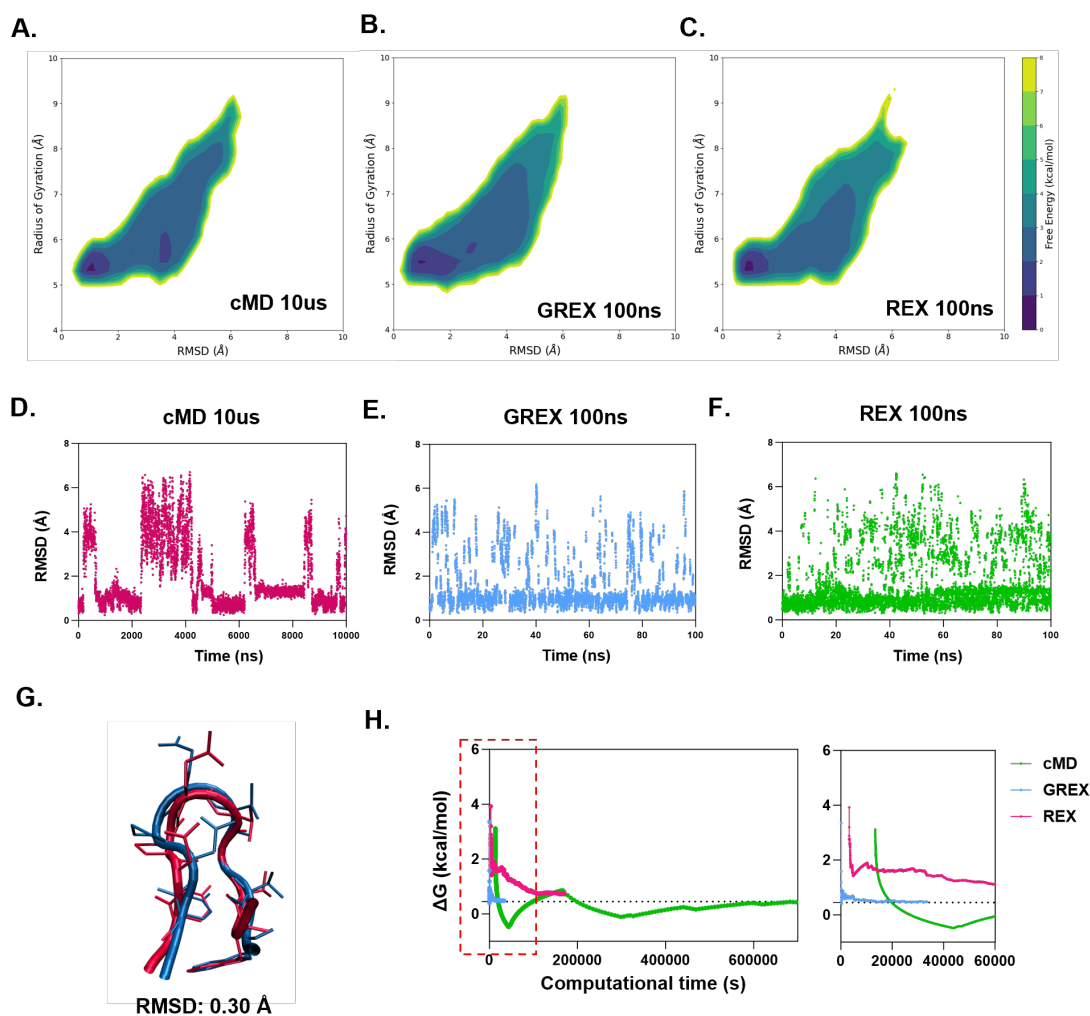

Figure 5. Evaluation of the chignolin model. (A-C) Two-dimensional FES as functions of RMSD and Rg from 10 μs cMD (F), 100 ns GREX (G), and 100 ns REX (H) simulations. (D-F) Backbone CA RMSDs of cMD (red), GREX (blue), and REX (green). (G) Starting from an unfolded initial structure, GREX reached a minimum RMSD of 0.30 nm, corresponding to native conformations (blue). (H) Convergence of the calculated $\Delta G$ as a function of simulation time. The dashed line indicates the experimental reference.

Beyond global folding thermodynamics, we also examined local conformational features through the FES of backbone dihedral angles ($\Phi_2$, $\Psi_2$) at the central glutamate residue, following previous enhanced sampling studies[41]. The 10 μs cMD reference simulation identified three local minima in this landscape (Figure 6A): two dominant minima at approximately ($\Phi_2 \approx -70°$, $\Psi_2 \approx 140°$) and ($\Phi_2 \approx -150°$, $\Psi_2 \approx 140°$), and a third minimum at ($\Phi_2 \approx -70°$, $\Psi_2 \approx -30°$) separated by a free energy barrier of ~4 kcal/mol. This substantial barrier prevents the metastable basin from being sampled in a 100 ns cMD simulation (Figure 6B). Both GREX and REX were able to capture this metastable basin within 100 ns (Figures 6C and 6D), with similar behavior observed for FESs of other backbone dihedrals (Figure S4). Importantly, GREX achieved this with only 34000 s (50 ns training + 100 ns production), representing a 5-fold speedup over REX (170000 s for 24 replicas × 100 ns) and an 18-fold speedup relative to the 10 μs cMD reference.

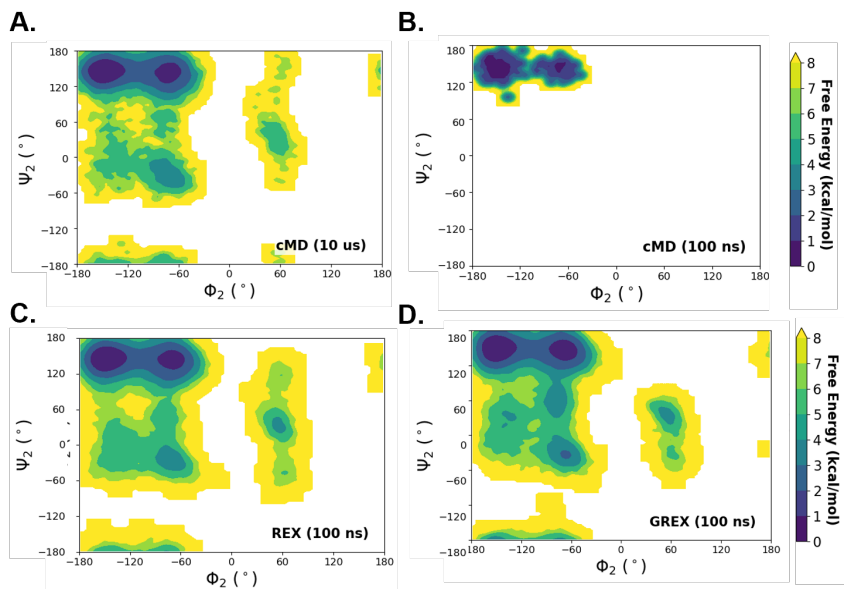

Figure 6. Two-dimensional FES of chignolin as functions of backbone dihedrals ($\Phi_2$ and $\Psi_2$) computed from (A) 10 μs cMD, (B) 100 ns cMD, (C) 100 ns REX, and (D) 100 ns GREX simulations.

## 4. Discussion and Conclusions

In this study, we introduced GREX, an enhanced sampling framework that integrates normalizing flows with the replica exchange framework to eliminate the need for a dense temperature ladder. Instead of relying on the conventional online relay of configurations through intermediate replicas, GREX employs a pair of trained normalizing flows: a Generator Flow that learns the high-temperature distribution from short MD data, and a Converter Flow that maps high-temperature configurations directly to the target temperature using the potential energy function. Generated configurations are introduced into the simulation through Metropolis exchange attempts, which correct any inaccuracies in the learned transformations and ensure that the resulting ensemble remains from the target Boltzmann distribution. Validation across three benchmark systems of increasing complexity, from a two-dimensional model potential to the 10-residue mini-protein chignolin, demonstrates that GREX is both thermodynamically rigorous and broadly effective, achieving 5- to 10-fold faster convergence speedups relative to conventional REX.

Although GREX builds on ideas from reservoir-based REX and normalizing flow methods, it addresses key limitations of both. Reservoir REX[15] pre-generates a physical ensemble at high temperature and substitutes it for the highest-temperature running replica, but still requires a full temperature ladder of N−1 intermediate replicas to connect the reservoir to the target temperature. GREX advances this concept further by replacing the fixed reservoir with a Generator Flow that learns the high-temperature distribution and produces new configurations on demand, while a Converter Flow maps them directly to the target temperature. This eliminates the intermediate ladder entirely, reducing the simulation to a single replica at the target temperature. Boltzmann Generators[21] use normalizing flows to bridge between distributions, but they are trained on target-temperature data and rely on importance reweighting to ensure thermodynamic rigor. For complex biomolecular systems, however, obtaining sufficient low-temperature training data is itself the central sampling challenge, and a model trained on limited low-temperature configurations risks inheriting the very sampling gaps it is intended to overcome. GREX circumvents this limitation by training on the high-temperature ensemble, which is readily easily explored in short MD simulations, and by using the potential energy function rather than target-temperature samples to guide the Converter Flow. The learned replica exchange (LREX) method[24] shares with GREX both the use of normalizing flows and a Metropolis criterion to ensure thermodynamic rigor. However, it requires parallel high-temperature simulations to run concurrently throughout the production stage, with the flow updated iteratively as new samples accumulate. GREX, in contrast, fully decouples training from production, so that a single offline training stage suffices for arbitrarily long production runs without additional high-temperature replicas.

Two limitations of the current implementation of GREX should be noted. First, the performance of GREX depends on adequate sampling of relevant conformational states in the high-temperature simulations. Configurations that are not sampled at high temperature cannot be learned by the generative model and will therefore be absent in the resulting ensemble at the target temperature, as illustrated by the missing conformational basin observed when only 0.01 ns of training data was used (Figure 4D). Second, GREX shared a limitation common to other normalizing flow-based methods: the scalability of normalizing flows to larger and more complex systems remains an open challenge[24]. Although GREX has been successfully applied here to systems with explicit solvent, extending the framework to significantly larger complex biomolecular systems may require deeper and more expressive flow architectures. Such models are typically more demanding to train and may introduce additional computational overhead. Encouragingly, normalizing flow methods are advancing rapidly, and improvements in flow architectures or training efficiency can be readily incorporated into GREX without altering the underlying exchange framework.

Looking ahead, key extensions of GREX include adaptive strategies to detect and address gaps in high-temperature sampling, as well as transfer learning across structurally related systems to reduce per-system training costs. Integration with CV-based methods also offers a promising direction, with GREX providing broad conformational diversity while CV-guided simulations refine free energy estimates along specific reaction coordinates. In summary, GREX provides a practical route to efficient enhanced sampling by replacing the temperature ladder with learned normalizing flows, achieving accuracy comparable to conventional REX at a fraction of the computational cost, with advantages that grow as system complexity increases.

NOTES

The code used in this work is publicly available at https://github.com/Hhuangsj/GREX. The normalizing flows have been implemented using the library https://github.com/noegroup/bgflow.

ACKNOWLEDGMENT

The authors gratefully acknowledge the financial support from Guangdong Basic and Applied Basic Research Foundation (2025A1515012114), Open Project of State Key Laboratory of Respiratory Disease (SKLRD-OP-202506), and Changjiang Scholars Award Program of Ministry of Education. S.H. would like to acknowledge the supported by the Excellent Graduate Student Cultivation Program of Jinan University (2025CXY353) and thank Dr. Michele Invernizzi for his helpful discussions on LREX

REFERENCE

(1) Karplus, M.; McCammon, J. A. Molecular Dynamics Simulations of Biomolecules. *Nat Struct Mol Biol* **2002**, *9* (9), 646–652.

(2) Lindorff-Larsen, K.; Piana, S.; Dror, R. O.; Shaw, D. E. How Fast-Folding Proteins Fold. *Science* **2011**, *334* (6055), 517–520.

(3) Liu, Y.; Tan, J.; Hu, S.; Hussain, M.; Qiao, C.; Tu, Y.; Lu, X.; Zhou, Y. Dynamics Playing a Key Role in the Covalent Binding of Inhibitors to Focal Adhesion Kinase. *J. Chem. Inf. Model.* **2024**, *64* (15), 6053–6061.

(4) Laio, A.; Parrinello, M. Escaping Free-Energy Minima. *Proceedings of the National Academy of Sciences* **2002**, *99* (20), 12562–12566.

(5) Ray, D.; Parrinello, M. Kinetics from Metadynamics: Principles, Applications, and Outlook. *J. Chem. Theory Comput.* **2023**, *19* (17), 5649–5670.

(6) Torrie, G. M.; Valleau, J. P. Nonphysical Sampling Distributions in Monte Carlo Free-Energy Estimation: Umbrella Sampling. *Journal of Computational Physics* **1977**, *23* (2), 187–199.

(7) Pietrucci, F. Strategies for the Exploration of Free Energy Landscapes: Unity in Diversity and Challenges Ahead. *Reviews in Physics* **2017**, *2*, 32–45.

(8) Earl, D. J.; Deem, M. W. Parallel Tempering: Theory, Applications, and New Perspectives. *Phys. Chem. Chem. Phys.* **2005**, *7* (23), 3910–3916.

(9) Swendsen, R. H.; Wang, J.-S. Replica Monte Carlo Simulation of Spin-Glasses. *Phys. Rev. Lett.* **1986**, *57* (21), 2607–2609.

(10) Sugita, Y.; Kitao, A.; Okamoto, Y. Multidimensional Replica-Exchange Method for Free-Energy Calculations. *The Journal of Chemical Physics* **2000**, *113* (15), 6042–6051.

(11) Fukunishi, H.; Watanabe, O.; Takada, S. On the Hamiltonian Replica Exchange Method for Efficient Sampling of Biomolecular Systems: Application to Protein Structure Prediction. *The Journal of Chemical Physics* **2002**, *116* (20), 9058–9067.

(12) Liu, P.; Kim, B.; Friesner, R. A.; Berne, B. J. Replica Exchange with Solute Tempering: A Method for Sampling Biological Systems in Explicit Water. *Proceedings of the National Academy of Sciences* **2005**, *102* (39), 13749–13754.

(13) Kasavajhala, K.; Simmerling, C. Exploring the Transferability of Replica Exchange Structure Reservoirs to Accelerate Generation of Ensembles for Alternate Hamiltonians or Protein Mutations. *J. Chem. Theory Comput.* **2023**, *19* (6), 1931–1944.

(14) Lyman, E.; Ytreberg, F. M.; Zuckerman, D. M. Resolution Exchange Simulation. *Phys. Rev. Lett.* **2006**, *96* (2), 028105.

(15) Okur, A.; Roe, D. R.; Cui, G.; Hornak, V.; Simmerling, C. Improving Convergence of Replica-Exchange Simulations through Coupling to a High-Temperature Structure Reservoir. *J. Chem. Theory Comput.* **2007**, *3* (2), 557–568.

(16) Ruscio, J. Z.; Fawzi, N. L.; Head-Gordon, T. How Hot? Systematic Convergence of the Replica Exchange Method Using Multiple Reservoirs.

(17) Rezende, D. J.; Mohamed, S. Variational Inference with Normalizing Flows. arXiv June 14, 2016.

(18) Kingma, D. P.; Dhariwal, P. Glow: Generative Flow with Invertible 1x1 Convolutions. arXiv July 10, 2018.

(19) Grathwohl, W.; Chen, R. T. Q.; Bettencourt, J.; Sutskever, I.; Duvenaud, D. FFJORD: Free-Form Continuous Dynamics for Scalable Reversible Generative Models. arXiv October 22, 2018.

(20) Dinh, L.; Sohl-Dickstein, J.; Bengio, S. Density Estimation Using Real NVP. arXiv February 27, 2017.

(21) Noé, F.; Olsson, S.; Köhler, J.; Wu, H. Boltzmann Generators: Sampling Equilibrium States of Many-Body Systems with Deep Learning. *Science* **2019**, *365* (6457), eaaw1147.

(22) Tabak, E. G.; Vanden-Eijnden, E. Density Estimation by Dual Ascent of the Log-Likelihood. *Comm. Math. Sci. 8* (1), 217–233.

(23) Dinh, L.; Krueger, D.; Bengio, Y. NICE: Non-Linear Independent Components Estimation. arXiv April 10, 2015.

(24) Invernizzi, M.; Kraemer, A.; Clementi, C.; Noe, F. Skipping the Replica Exchange Ladder with Normalizing Flows. *J. Phys. Chem. Lett.* **2022**.

(25) Case, D. A.; Aktulga, H. M.; Belfon, K.; Cerutti, D. S.; Cisneros, G. A.; Cruzeiro, V. W. D.; Forouzesh, N.; Giese, T. J.; Götz, A. W.; Gohlke, H.; Izadi, S.; Kasavajhala, K.; Kaymak, M. C.; King, E.; Kurtzman, T.; Lee, T.-S.; Li, P.; Liu, J.; Luchko, T.; Luo, R.; Manathunga, M.; Machado, M. R.; Nguyen, H. M.; O'Hearn, K. A.; Onufriev, A. V.; Pan, F.; Pantano, S.; Qi, R.; Rahnamoun, A.; Risheh, A.; Schott-Verdugo, S.; Shajan, A.; Swails, J.; Wang, J.; Wei, H.; Wu, X.; Wu, Y.; Zhang, S.; Zhao, S.; Zhu, Q.; Cheatham, T. E. I.; Roe, D. R.; Roitberg, A.; Simmerling, C.; York, D. M.; Nagan, M. C.; Merz, K. M. Jr. AmberTools. *J. Chem. Inf. Model.* **2023**, *63* (20), 6183–6191.

(26) Berman, H. M.; Westbrook, J.; Feng, Z.; Gilliland, G.; Bhat, T. N.; Weissig, H.; Shindyalov, I. N.; Bourne, P. E. The Protein Data Bank. *Nucleic Acids Research* **2000**, *28* (1), 235–242.

(27) Eastman, P.; Galvelis, R.; Peláez, R. P.; Abreu, C. R. A.; Farr, S. E.; Gallicchio, E.; Gorenko, A.; Henry, M. M.; Hu, F.; Huang, J.; Krämer, A.; Michel, J.; Mitchell, J. A.; Pande, V. S.; Rodrigues, J. P.; Rodriguez-Guerra, J.; Simmonett, A. C.; Singh, S.; Swails, J.; Turner, P.; Wang, Y.; Zhang, I.; Chodera, J. D.; De Fabritiis, G.; Markland, T. E. OpenMM 8: Molecular Dynamics Simulation with Machine Learning Potentials. *J. Phys. Chem. B* **2024**, *128* (1), 109–116.

(28) Maier, J. A.; Martinez, C.; Kasavajhala, K.; Wickstrom, L.; Hauser, K. E.; Simmerling, C. ff14SB: Improving the Accuracy of Protein Side Chain and Backbone Parameters from ff99SB. *J. Chem. Theory Comput.* **2015**, *11* (8), 3696–3713.

(29) Jorgensen, W. L.; Chandrasekhar, J.; Madura, J. D.; Impey, R. W.; Klein, M. L. Comparison of Simple Potential Functions for Simulating Liquid Water. *The Journal of Chemical Physics* **1983**, *79* (2), 926–935.

(30) Harvey, M. J.; De Fabritiis, G. An Implementation of the Smooth Particle Mesh Ewald Method on GPU Hardware. *J. Chem. Theory Comput.* **2009**, *5* (9), 2371–2377.

(31) Zhang, Z.; Liu, X.; Yan, K.; Tuckerman, M. E.; Liu, J. Unified Efficient Thermostat Scheme for the Canonical Ensemble with Holonomic or Isokinetic Constraints via Molecular Dynamics. *J. Phys. Chem. A* **2019**, *123* (28), 6056–6079.

(32) McGibbon, R. T.; Beauchamp, K. A.; Harrigan, M. P.; Klein, C.; Swails, J. M.; Hernández, C. X.; Schwantes, C. R.; Wang, L.-P.; Lane, T. J.; Pande, V. S. MDTraj: A Modern Open Library for the Analysis of Molecular Dynamics Trajectories. *Biophysical Journal* **2015**, *109* (8), 1528–1532.

(33) Invernizzi, M.; Parrinello, M. Making the Best of a Bad Situation: A Multiscale Approach to Free Energy Calculation. *J. Chem. Theory Comput.* **2019**, *15* (4), 2187–2194.

(34) Biswas, M.; Lickert, B.; Stock, G. Metadynamics Enhanced Markov Modeling of Protein Dynamics. *J. Phys. Chem. B* **2018**, *122* (21), 5508–5514.

(35) Botan, V.; Backus, E. H. G.; Pfister, R.; Moretto, A.; Crisma, M.; Toniolo, C.; Nguyen, P. H.; Stock, G.; Hamm, P. Energy Transport in Peptide Helices. *Proceedings of the National Academy of Sciences* **2007**, *104* (31), 12749–12754.

(36) Lindorff-Larsen, K.; Piana, S.; Dror, R. O.; Shaw, D. E. How Fast-Folding Proteins Fold. *Science* **2011**, *334* (6055), 517–520.

(37) Honda, S.; Akiba, T.; Kato, Y. S.; Sawada, Y.; Sekijima, M.; Ishimura, M.; Ooishi, A.; Watanabe, H.; Odahara, T.; Harata, K. Crystal Structure of a Ten-Amino Acid Protein. *J. Am. Chem. Soc.* **2008**, *130* (46), 15327–15331.

(38) Miao, Y.; Feixas, F.; Eun, C.; McCammon, J. A. Accelerated Molecular Dynamics Simulations of Protein Folding. *Journal of Computational Chemistry* **2015**, *36* (20), 1536–1549.

(39) Shaffer, P.; Valsson, O.; Parrinello, M. Enhanced, Targeted Sampling of High-Dimensional Free-Energy Landscapes Using Variationally Enhanced Sampling, with an Application to Chignolin. *Proc. Natl. Acad. Sci. U.S.A.* **2016**, *113* (5), 1150–1155.

(40) Okumura, H. Temperature and Pressure Denaturation of Chignolin: Folding and Unfolding Simulation by Multibaric-multithermal Molecular Dynamics Method. *Proteins* **2012**, *80* (10), 2397–2416.

(41) Shaffer, P.; Valsson, O.; Parrinello, M. Enhanced, Targeted Sampling of High-Dimensional Free-Energy Landscapes Using Variationally Enhanced Sampling, with an Application to Chignolin. *Proceedings of the National Academy of Sciences* **2016**, *113* (5), 1150–1155.

# Generative Replica-Exchange: A Flow-based Framework for Accelerating Replica Exchange Simulations

## Contents

**Double-well potential**

The N-dimensional double-well potential introduced in Michele's etc work[1], shown in Figure 2. The system with a single particle moving with a Langevin dynamics with two minima. The translations between them are rare in low temperature. We define the first two dimensions as coordinates x1 and x2, feel the double-well potential, while all the other are subject to a harmonic potential:

$$U_{dw}(x) = U_{wq}(x_1, x_2) + \frac{15}{2} \sum_{i=3}^{N} x_i^2$$

Where $U_{wq}(x)$ is the modified Wolfe-Quapp potential[]:

$$U_{wq}(x_1, x_2) = 2(y_1^4 + y_2^4 - 2y_1^2 - 4y_2^2 + 2y_1 y_2 + 0.8y_1 + 0.1y_2 + 9.28)$$

And $y_{[1,2]} = y_{[1,2]}(x_1, x_2)$ are rotated coordinates,

$$y_1(x_1, x_2) = x_1 \cos(\theta) - x_2 \sin(\theta), where\ \theta = 0.15\pi$$
$$y_2(x_1, x_2) = x_1 \sin(\theta) + x_2 \cos(\theta), where\ \theta = 0.15\pi$$

**List of Figures**

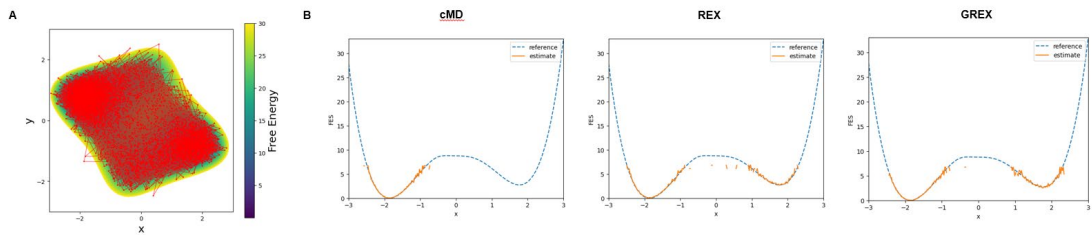

**Figure S1**. (A) The free energy surface and representative trajectories obtained from cMD at high temperature (T = 5). (B) The one-dimensional marginal probability distributions along the x-coordinate.

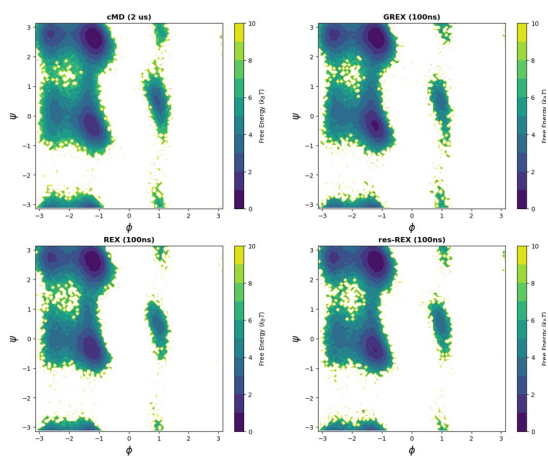

**Figure S2**. Free energy surface at 300 K as a function of $\Phi$ and $\Psi$, obtained from Cmd (2 μs), GREX (100 ns), REX (100 ns), and res-REX (100 ns).

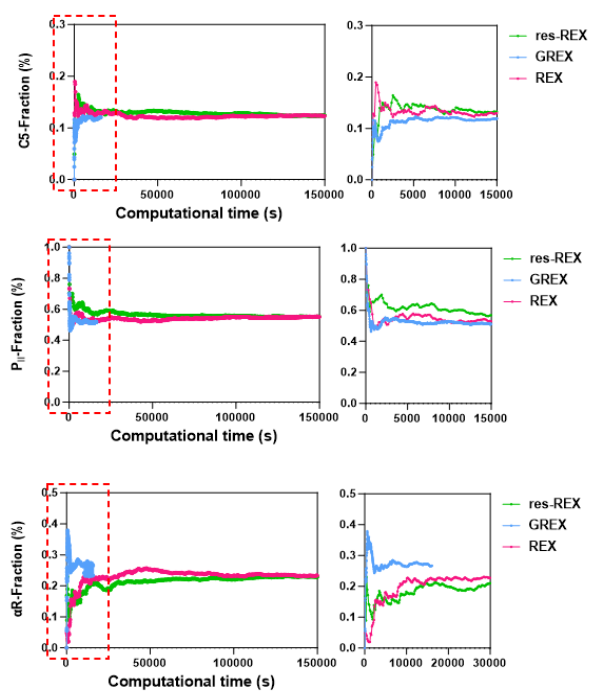

**Figure S3**. Convergence of the {C5, $P_{II}$, and $\alpha_R$} basin population as a function of computational time for GREX, res-REX, and conventional REX. The dashed box highlights the early-time regime.

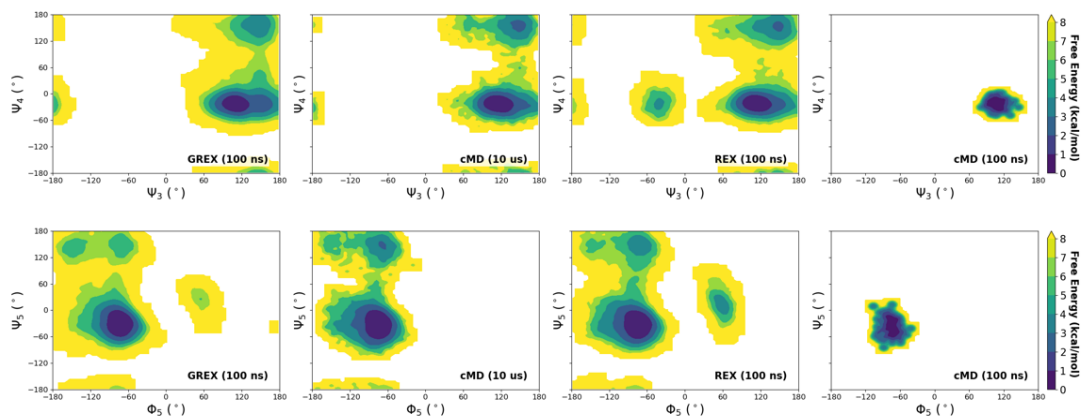

**Figure S4**. Two-dimensional FES as functions of backbone dihedrals ($\Psi_3$, $\Psi_4$) and ($\Phi_5$, $\Psi_5$) in chignolin[3] from (A) 10 μs cMD, (B) 100 ns cMD, (C) 100 ns REX, and (D) 100 ns GREX.

**Reference**

(1) Invernizzi, M.; Parrinello, M. Making the Best of a Bad Situation: A Multiscale Approach to Free Energy Calculation. *J. Chem. Theory Comput.* **2019**, *15* (4), 2187–2194.
(2) Dinh, L.; Sohl-Dickstein, J.; Bengio, S. Density Estimation Using Real NVP. arXiv February 27, 2017.
(3) Shaffer, P.; Valsson, O.; Parrinello, M. Enhanced, Targeted Sampling of High-Dimensional Free-Energy Landscapes Using Variationally Enhanced Sampling, with an Application to Chignolin. *Proceedings of the National Academy of Sciences* **2016**, *113* (5), 1150–1155.